\documentclass[useAMS,usenatbib]{mn2e}

\usepackage{graphicx}
\usepackage{amsmath}
\usepackage{amssymb}
\usepackage{dcolumn}
\newcommand{\mc}[1]{\multicolumn{1}{c}{#1}}

\title[Discovery of the first $\tau$\,Sco analogues: HD\,66665 and HD\,63425]{Discovery of the first $\tau$\,Sco analogues: HD\,66665 and HD\,63425}

\author[Petit et al.]{V. Petit$^{1}$, 
D.~L. Massa$^{2}$,
W.~L.~F. Marcolino$^{3}$,
G.~A. Wade$^{4}$,
R. Ignace $^{5}$   \newauthor  and
The~MiMeS~Collaboration
\thanks{Based on observations obtained at the Canada-France-Hawaii Telescope (CFHT) which is operated by the National Research Concil of Canada, the Institut National des Sciences de l'Univers of the Centre National de la Recherche Scientifique of France, and the University of Hawaii.}\\
$^{1}$ Dept. of Geology \& Astronomy, West Chester University, West Chester, PA 19383, E-mail: VPetit@wcupa.edu\\
$^{2}$ Space Telescope Science Institute, 3700 N. San Martin Drive, Baltimore, MD 21218, USA \\
$^{3}$ Observat\'orio Nacional-MCT, Rua Jos\'e Cristino, 77, CEP 20921-400, S\~ao Crist\'ov\~ao, Rio de Janeiro, Brasil\\
$^{4}$ Dept. of Physics, Royal Military College of Canada, PO Box 17000, Stn Forces, Kingston, Canada, K7K 4B4\\
$^{5}$ Department of Physics \& Astronomy, East Tennessee State University, Box 70652, Johnson City, TN 37614, USA}

\begin{document}
%
%
%


\def\jnl@style{\it}
\def\aaref@jnl#1{{\jnl@style#1}}

\def\aaref@jnl#1{{\jnl@style#1}}

\def\aj{\aaref@jnl{AJ}}                   
\def\araa{\aaref@jnl{ARA\&A}}             
\def\apj{\aaref@jnl{ApJ}}                 
\def\apjl{\aaref@jnl{ApJ}}                
\def\apjs{\aaref@jnl{ApJS}}               
\def\ao{\aaref@jnl{Appl.~Opt.}}           
\def\apss{\aaref@jnl{Ap\&SS}}             
\def\aap{\aaref@jnl{A\&A}}                
\def\aapr{\aaref@jnl{A\&A~Rev.}}          
\def\aaps{\aaref@jnl{A\&AS}}              
\def\azh{\aaref@jnl{AZh}}                 
\def\baas{\aaref@jnl{BAAS}}               
\def\jrasc{\aaref@jnl{JRASC}}             
\def\memras{\aaref@jnl{MmRAS}}            
\def\mnras{\aaref@jnl{MNRAS}}             
\def\pra{\aaref@jnl{Phys.~Rev.~A}}        
\def\prb{\aaref@jnl{Phys.~Rev.~B}}        
\def\prc{\aaref@jnl{Phys.~Rev.~C}}        
\def\prd{\aaref@jnl{Phys.~Rev.~D}}        
\def\pre{\aaref@jnl{Phys.~Rev.~E}}        
\def\prl{\aaref@jnl{Phys.~Rev.~Lett.}}    
\def\pasp{\aaref@jnl{PASP}}               
\def\pasj{\aaref@jnl{PASJ}}               
\def\qjras{\aaref@jnl{QJRAS}}             
\def\skytel{\aaref@jnl{S\&T}}             
\def\solphys{\aaref@jnl{Sol.~Phys.}}      
\def\sovast{\aaref@jnl{Soviet~Ast.}}      
\def\ssr{\aaref@jnl{Space~Sci.~Rev.}}     
\def\zap{\aaref@jnl{ZAp}}                 
\def\nat{\aaref@jnl{Nature}}              
\def\iaucirc{\aaref@jnl{IAU~Circ.}}       
\def\aplett{\aaref@jnl{Astrophys.~Lett.}} 
\def\apspr{\aaref@jnl{Astrophys.~Space~Phys.~Res.}}
\def\bain{\aaref@jnl{Bull.~Astron.~Inst.~Netherlands}} 
\def\fcp{\aaref@jnl{Fund.~Cosmic~Phys.}}  
\def\gca{\aaref@jnl{Geochim.~Cosmochim.~Acta}}   
\def\grl{\aaref@jnl{Geophys.~Res.~Lett.}} 
\def\jcp{\aaref@jnl{J.~Chem.~Phys.}}      
\def\jgr{\aaref@jnl{J.~Geophys.~Res.}}    
\def\jqsrt{\aaref@jnl{J.~Quant.~Spec.~Radiat.~Transf.}}
\def\memsai{\aaref@jnl{Mem.~Soc.~Astron.~Italiana}}
\def\nphysa{\aaref@jnl{Nucl.~Phys.~A}}   
\def\physrep{\aaref@jnl{Phys.~Rep.}}   
\def\physscr{\aaref@jnl{Phys.~Scr}}   
\def\planss{\aaref@jnl{Planet.~Space~Sci.}}   
\def\procspie{\aaref@jnl{Proc.~SPIE}}   

\let\astap=\aap
\let\apjlett=\apjl
\let\apjsupp=\apjs
\let\applopt=\ao

\date{Accepted 2010 December 17. Received 2010 December 16; in original form 2010 October 30}

\pagerange{\pageref{firstpage}--\pageref{lastpage}} \pubyear{2010}

\maketitle

\label{firstpage}

\begin{abstract}
The B0.2 V magnetic star $\tau$\,Sco stands out from the larger population of massive OB stars due to its high X-ray activity, peculiar wind diagnostics and highly complex magnetic field. This paper presents the discovery of the first two $\tau$\,Sco analogues - HD\,66665 and HD\,63425, identified by the striking similarity of their UV spectra to that of $\tau$~Sco.
ESPaDOnS spectropolarimetric observations were secured by the Magnetism in Massive Stars CFHT Large Program, in order to characterize the stellar and magnetic properties of these stars.
\textsc{cmfgen} modelling of optical ESPaDOnS spectra and archived IUE UV spectra showed that these stars have stellar parameters similar to those of $\tau$\,Sco. 
A magnetic field of similar surface strength is found on both stars, reinforcing the connection between the presence of a magnetic field and wind peculiarities. 
However, additional phase-resolved observations will be required in order to assess the potential complexity of the magnetic fields, and verify if the wind anomalies are linked to this property.
\end{abstract}


\section{The magnetic B-type star $\tau$~Sco}

Very little is known about the magnetic fields of hot, massive OB stars. Due at least in part to the challenges of measurement, direct evidence of magnetic fields in massive stars is remarkably rare - there are perhaps 30 known examples - and few have been studied in detail.

Even as a member of the elusive class of magnetic massive stars, the B0.2\,V star $\tau$\,Sco is recognized to be a peculiar and outstanding object.
First, $\tau$~Sco is distinguished by its magnetic field, which is unique because it is structurally far more complex than the mostly-dipolar (multipole mode $l=1$) fields usually observed in magnetic OB stars.  According to the Zeeman Doppler Imaging performed by \citet{2006MNRAS.370..629D}, the magnetic topology of $\tau$\,Sco presents significant power in multipole modes up to $l=5$ with a mean surface field strength of $\sim300$\,G, and the extrapolated 3D field structure shows an intricate assortment of open field lines and closed magnetic loops \citep[illustrated in Fig. 11 and 12 of][]{2006MNRAS.370..629D}. 

\begin{figure*}
\includegraphics[width=87mm]{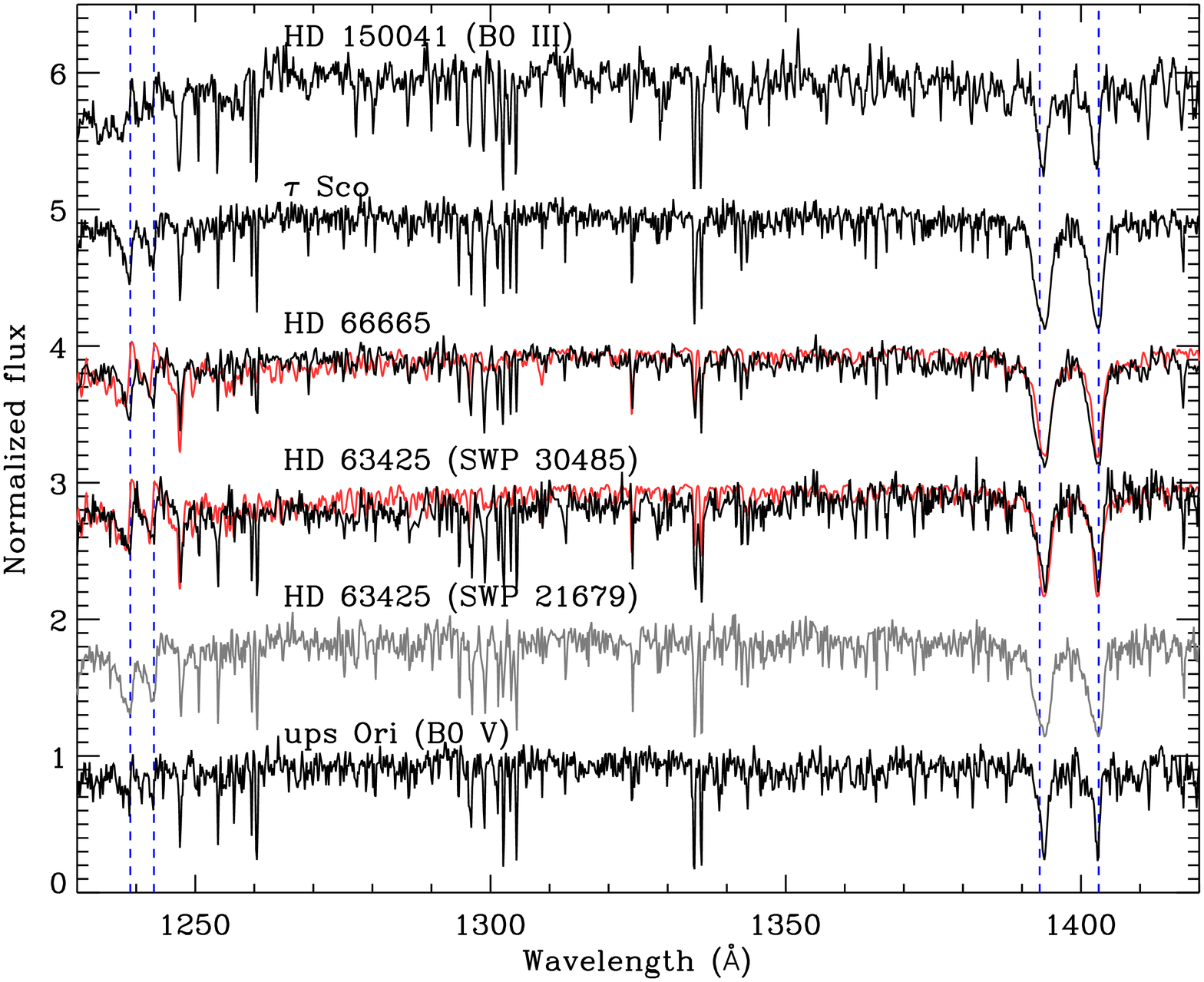}
\includegraphics[width=87mm]{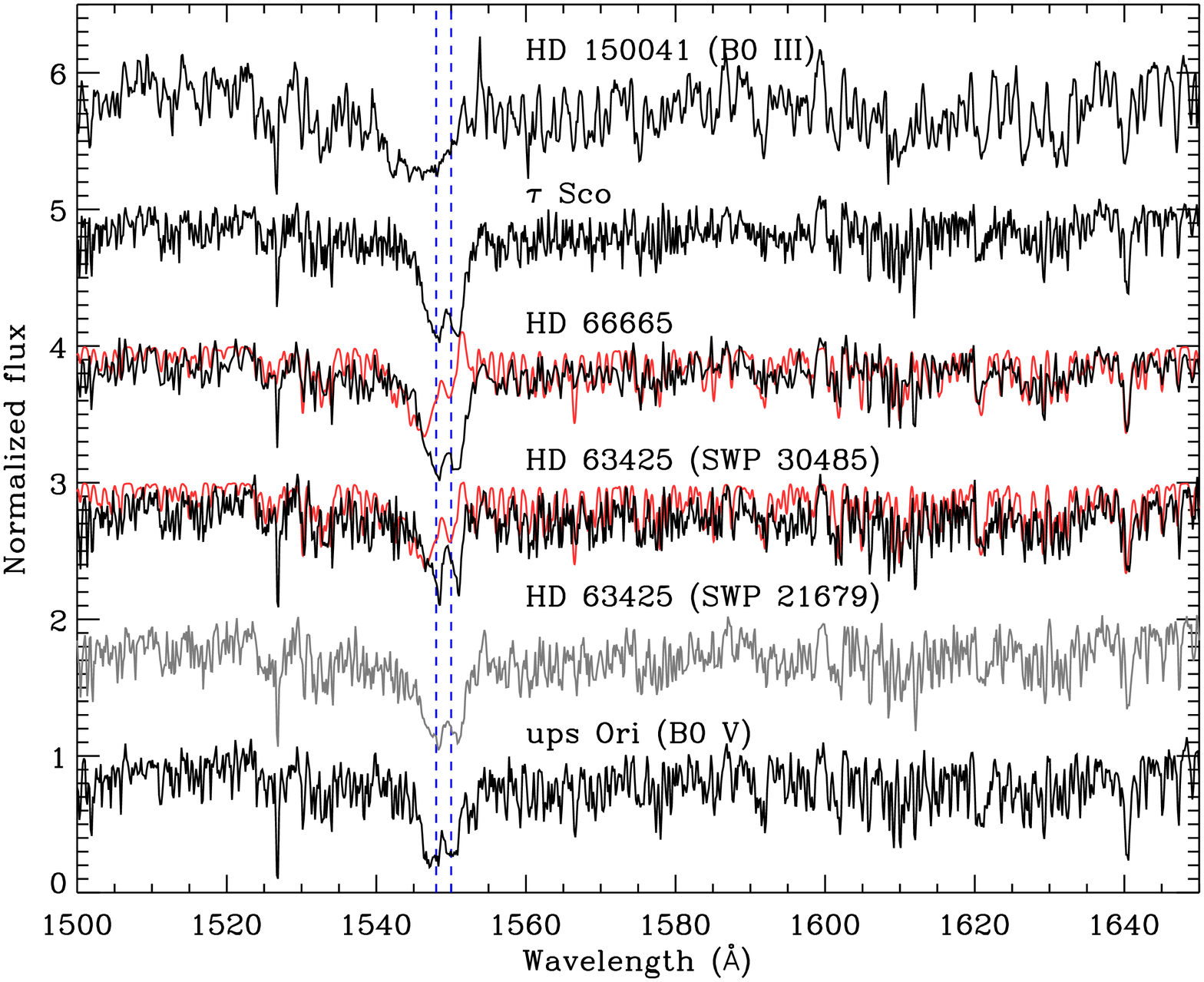}
\caption{\label{fig|uv} IUE spectra of $\tau$\,Sco and its analogues HD\,66665 and HD\,63425 (second to fifth spectra). For comparison, typical spectra of a B0 dwarf and a B0 giant are shown (bottom and top respectively). The dashed lines indicate the wind lines N\,\textsc{v}\,$\lambda \lambda 1239, 1243$, Si\,\textsc{iv}\,$\lambda\lambda 1393, 1403$ and C\,\textsc{iv}\,$\lambda \lambda 1548, 1550$. The best-fit synthetic spectra, from which stellar parameters were derived from for HD\,66665 and HD\,63425, are shown in red. }
\end{figure*}

Secondly, $\tau$\,Sco also stands out from the crowd of early-B stars because of its stellar wind anomalies, as diagnosed through its odd UV spectrum and X-ray emission. 
The optical and non-resonance UV lines of $\tau$~Sco, that probe the conditions of the photosphere, are typical of an early-B dwarf star. 
However, the UV resonance lines arising from the stellar wind are not those typical of such a star. 
Figure \ref{fig|uv} shows a comparison of the UV spectrum of $\tau$\,Sco (second spectrum) with the spectrum of a normal B0 main sequence star (bottom spectrum) and a B0 giant (top spectrum). The resonance wind lines of C\,\textsc{iv} (right panel), Si\,\textsc{iv} and N\,\textsc{v} (left panel) are indicated with dashed lines.
The UV wind line morphology of normal, early B stars conforms to a 2-D spectral grid \citep{1995NASRP1363.....W}.  Typically,  C\,\textsc{iv} strengthens with increasing temperature and luminosity. N\,\textsc{v} is very weak on the main sequence at spectral type B0\,V but strengthens with temperature and luminosity.  
For stars with fixed C\,\textsc{iv} and N\,\textsc{v} profiles, Si\,\textsc{iv} is strictly luminosity dependent and breaks the degeneracy. $\tau$\,Sco does not fit into this grid. It has strong N\,\textsc{v} indicating that it should be well above the main
sequence.  However, its C\,\textsc{iv} lines are only slightly stronger than typical and not distinctly wind-like, suggesting a near main sequence luminosity.  Finally, its Si\,\textsc{iv} profiles are unique, a bit stronger than typical class V stars, but unlike normal, early giants.  As a result, this stars lie outside of the normal classification grid, suggesting a more highly ionized outflow than typical. The hard X-ray emission of $\tau$\,Sco also suggests hot plasma, in excess of 10\,MK \citep{1994ApJ...421..705C}. We suspect that this is connected to the strange UV properties. Like its magnetic field, the stellar wind of $\tau$\,Sco possesses observational characteristics that make it unique among OB stars.

Interestingly, the wind lines of $\tau$~Sco have been shown to vary periodically with the star's 41~d rotation period \citep{2006MNRAS.370..629D}, which is interpreted to mean that the wind is structured and modulated by the magnetic field. Clearly the magnetic field exerts an important influence on the wind dynamics. What is not clear is whether the wind-line anomalies described above are a consequence of the unusual complexity of $\tau$ Sco's magnetic field, a general consequence of wind confinement in this class of star, or perhaps even unrelated to the presence of a magnetic field.  Because such wind anomalies have never been observed in any other star, magnetic or not, this issue has remained unresolved.

The identification and analysis of additional stars with wind properties similar to $\tau$~Sco would therefore represent an important step toward understanding the origin of these peculiarities. 
In this paper, we present two early B-type stars - HD\,66665 and HD\,63425 - that we identified to be the first $\tau$\,Sco analogues based on their UV spectra, which are strikingly similar to the UV spectrum of $\tau$\,Sco. Figure \ref{fig|uv} shows \text{IUE} archives spectra of these two stars (third to fifth spectra from the top).
The discovery of wind anomalies naturally led to an investigation of the magnetic properties of these stars by the Magnetism in Massive Stars (MiMeS) collaboration \citep{2010arXiv1009.3563W}.

\begin{table}
	\caption{\label{tab|obslog} Journal of ESPaDOnS observation listing the UT date, the heliocentric Julian date (2\,400\,000+), the total exposure time and the s/n per 1.8\,km\,s$^{-1}$ velocity bin at 540\,nm.
Column 5 gives the probability that the observed signal in Stokes V LSD profiles is due only to noise, and column 6 the global longitudinal field component obtained from the first moment of the Stokes V profiles. The available archival \textit{IUE} spectra are also listed.}
	\begin{tabular}{c c c c c D{*}{\,\pm\,}{3,3} }
	\hline
	Date & HJD & $t_\mathrm{exp}$ & s/n & FAP & \mc{$B_l~^1$} \\
	 (UT)  &     &  (s) & &  & \mc{(G)} \\
	\hline
	\multicolumn{6}{c}{\textbf{HD\,66665}} \\
	2010/02/26 & 55\,253.97 & $4\,400$ & 728 & $<10^{-8}$  & -93*4 \\
	2010/03/02 & 55\,257.98 & $1\,200$   & 236 & $<10^{-8}$ & -3*15 \\
	2010/03/04 & 55\,259.98 & $1\,200$   & 330 & $<10^{-8}$ & 9*10 \\
	2010/03/05 & 55\,260.98 & $1\,200$   & 276 & $2\times10^{-6}$ & 41*13 \\
	1983/12/04 & 45\,436.77 & 900 &  \multicolumn{3}{c}{IUE swp\,21680} \\
	1987/03/11 & 46\,866.49 & 420 &  \multicolumn{3}{c}{IUE swp\,30484} \\
	1987/03/12 & 46\,866.58 & 1\,080 & \multicolumn{3}{c}{IUE swp\,30486} \\
	\hline
	\multicolumn{6}{c}{\textbf{HD\,63425}} \\
	2010/02/28 & 55\,255.84 & $1\,200$   & 520 & $<10^{-8}$ & 117*8 \\
	2010/03/03 & 55\,258.88 & $1\,200$   & 289 & $<10^{-8}$ & 113*15 \\
	2010/03/07 & 55\,262.76 & $1\,200$   & 435 & $<10^{-8}$ & 132*10 \\
	2010/03/08 & 55\,263.83 & $1\,200$   & 361 & $<10^{-8}$ & 125*12 \\
	1987/03/12 & 46\,866.53 & 600 &  \multicolumn{3}{c}{IUE swp\,21679} \\
	1983/12/04 & 45\,672.74 & 720&  \multicolumn{3}{c}{IUE swp\,30485} \\
	\hline
	\end{tabular}
	$^1$ A detectable Stokes V signal can be observed with high-resolution instruments even if the global longitudinal field is null \citep[see][]{2010arXiv1010.2248P}. 
\end{table}

\begin{figure}
\includegraphics[width=84mm]{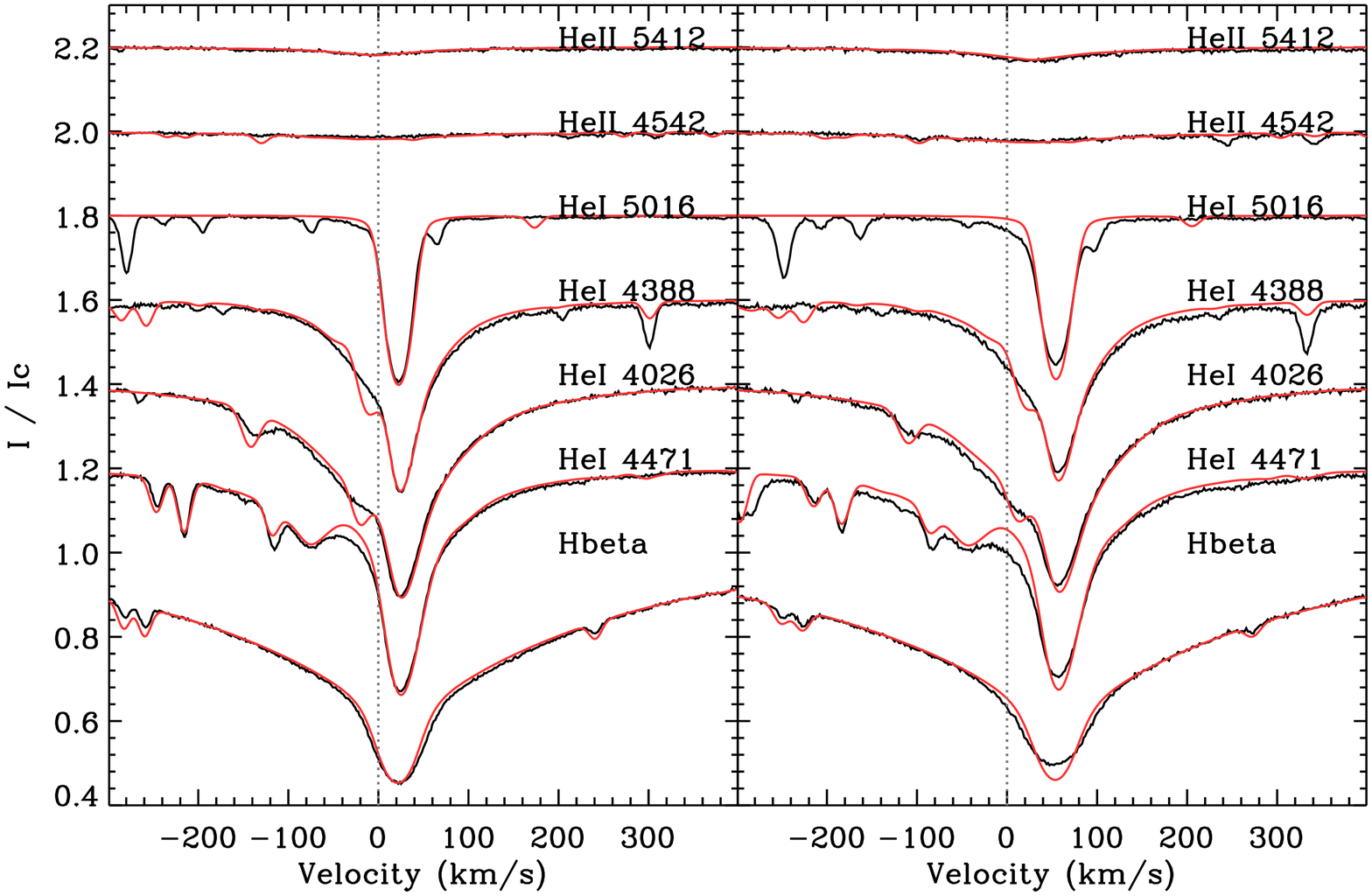}
\caption{\label{fig|vis} Fits to the optical He\,\textsc{i-ii} and H$\beta$ lines for HD\,66665 (left) and HD\,63425 (right) }
\end{figure} 

\begin{table}
	\caption{\label{tab|param} Summary of stellar properties of $\tau$\,Sco, HD\,66665 and HD\,63425. }
	\begin{tabular}{l c c c}
	\hline
	 & $\tau$\,Sco$^1$ & HD\,66665 & HD\,63425 \\
	 \hline
	 Spec. type & B0.2\,V & B0.5\,V & B0.5\,V\\
	 $T_\mathrm{eff}$  (kK)& $31\pm1$ & $28.5\pm1.0$ & $29.5\pm1.0$ \\
	 $\log g$  (cgs)& $4.0\pm0.1$ & $3.9\pm0.1$ & $4.0\pm0.1$ \\
	 $R_\star$ (R$_\odot$)  & $5.6\pm0.8$ & $5.5\stackrel{+3.3}{_{-2.7}}$ & $6.8\stackrel{+4.8}{_{-2.0}}$ \\
 	 $\log(L_\star/\mathrm{L}_\odot)$ & $4.47\pm0.13$ & $4.25\stackrel{+0.40}{_{-0.60}}$ & $4.50\stackrel{+0.46}{_{-0.30}}$ \\
	 $M_\star$ (M$_\odot$) & $11\pm4$ &  $9\stackrel{+5}{_{-7}}$ & $17\stackrel{+34}{_{-9}}$ \\
	 $v\sin i$ (km\,s$^{-1}$) & $<13$ & $\lesssim10$ & $\lesssim15$ \\
	 $\dot{M}$ ($10^{-9}$\,M$_\odot$\,yr$^{-1}$)  & $61\stackrel{+10}{_{-2}}$ &  $< 0.45$  & $< 0.75$  \\
	$v_\infty$  (km\,s$^{-1}$)&    $\sim2000$        & $\sim1400$ & $\sim1700$ \\
	\hline
	\end{tabular}
	
	\medskip
	$^1$ Paramters from \citet{2006A&A...448..351S}, and \citet{2005A&A...441..711M} for $\dot{M}$ and $v_\infty$.
\end{table}

\section{Observations}

Spectropolarimetric observations of HD\,66665 and HD\,63425 were obtained with ESPaDOnS at the Canada-France-Hawaii Telescope. We acquired 4 high-resolution ($R\sim65\,000$) broad-band (370-1050\,nm) intensity (Stokes I) and circular polarisation (Stokes V) spectra for each star, during about 10 consecutive nights in Feb/Mar 2010.
A summary of the ESPaDOnS observations is given in Table \ref{tab|obslog}.
For a complete description of observation procedure and reduction procedure with the \textsc{libre-esprit} package provided by CFHT, see \citet{1997MNRAS.291..658D}.
In addition to the ESPaDOnS observations, the available archival \textit{IUE} spectra are also listed in Table \ref{tab|obslog}.

\section{Stellar parameters}

In order to determine the stellar and wind parameters of HD\,66665 and HD\,63425 we used non-LTE model atmospheres and spectrum synthesis from the \textsc{cmfgen} code. \textsc{cmfgen} solves for the radiative transfer equation in a spherically symmetric outflow together with the radiative and statistical equilibrium equations, taking into account effects such as line-blanketing and X-rays \citep[for details, see][]{1998ApJ...496..407H}.

We used optical spectra to derive the surface gravity ($\log g$) and effective temperature ($T_\mathrm{eff}$), using classic line diagnostics (i.e., Balmer line wings and He\,\textsc{i-ii} lines). UV spectra from \textit{IUE} 
were used to infer stellar wind properties. All results are summarized in Table \ref{tab|param}, along with those of $\tau$\,Sco for comparison. A solar chemical composition was adopted, following \citet{2009ARA&A..47..481A}. We postpone a detailed analysis regarding chemical abundances for a future study. 

\subsection{HD\,66665}

The various estimates for the distance of HD\,66665 range from about 1 to 2\,kpc \citep{1985ApJS...59..397S,1987ApJ...314..380S,1994ApJS...93..211D,2007A&A...474..653V}. 
We adopted an intermediate value of 1.5\,kpc and assumed an uncertainty of a factor of two\footnote{Note that the recently revised parallax for HD\,66665 is very uncertain, namely, $\pi = -0.94 \pm1.05$\,mas \citep{2007A&A...474..653V}.}. The stellar luminosity $L_\star$ was then derived by fitting the spectral energy distribution using the \textit{IUE} spectrum and UBVJHK photometry, converted to flux points. An $E(B-V) = 0.02$ and $R = 3.1$ were used to fit the observed data \citep{1994ApJS...93..211D}. The stellar radius ($R_\star$) and mass ($M_\star$) follow directly from the chosen $L_\star$, $T_\mathrm{eff}$, and $\log g$.

A very good agreement is obtained for the optical diagnostic lines in HD\,66665. In Fig. \ref{fig|vis} (left) we present fits to different He\,\textsc{i-ii} lines and H$\beta$. Our model was convolved with a $v\sin i$ of 8\,km\,s$^{-1}$. We note however that this value is an estimate. Values in excess of 10\,km\,s$^{-1}$ tend to produce profiles which are generally broader than the ones observed but accurate lower values are difficult to determine. No significant variability was observed in the optical spectra of HD\,66665.

The UV spectrum of HD 66665 was first pointed out as unusual by \citet{1992BAAS...24..800G}. The absorption shown by C\,\textsc{iv} cannot be reproduced in detail by our models (see Figure \ref{fig|uv}). Several tests were performed, but the observed profile is always too deep compared to the synthetic ones. On the other hand, we could achieve an acceptable fit for the Si\,\textsc{iv}, and N\,\textsc{v} lines. The inclusion of X-rays in our \textsc{cmfgen} models was essential in producing the N\,\textsc{v} feature, since the wind structure is not sufficiently ionized without this component. 

Despite the problem found for C\,\textsc{iv}, we could derive an upper limit for the mass-loss rate ($\dot{M}$) from the emission part of the synthetic P-Cygni profile of this same line. When $\dot{M}$ is greater than about $4.5\times 10^{-10}$\,M$_\odot$\,yr$^{-1}$, the model predicts a relatively strong P-Cygni emission which is clearly absent in the observed spectrum, as it is in the spectrum of $\tau$\,Sco. Furthermore, only a rough estimate for the wind terminal velocity ($v_\infty\simeq1\,400$\,km\,s$^{-1}$) could be made.

\subsection{HD\,63425}

We adopted a distance to HD\,63425 of 1136\,pc \citep[from $\pi = 0.88 \pm 0.36$\,mas;][]{2007A&A...474..653V}.
The stellar luminosity was then derived in the same way as in HD\,66665.
For the reddening parameters, we use $R=3.1$ and $E(B-V)=0.10$ \citep{1987ApJ...314..380S}. 
Our fits to the He\,\textsc{i-ii} diagnostic lines and H$\beta$ are presented in Figure \ref{fig|vis} (right). The optical spectra of HD\,63425 and HD\,66665 are similar. However, our analysis indicates a slightly higher effective temperature and surface gravity for HD\,63425. Moreover, Balmer line cores in HD\,63425 are shallower, especially H$\alpha$ and H$\beta$. Line filling by a much higher mass-loss rate than the one in our final model is not a reasonable explanation here, since it would produce strong UV wind profiles which are not observed. A possible explanation for this line filling could be contamination (emission) from material trapped by the magnetic field, distinct from the outflowing wind. This problem does not occur in HD\,66665, where we have good fits to entire Balmer line profiles. 

The apparent rotational velocity of HD\,63425 is low, but an accurate value is difficult to infer from the observed spectrum. Values above 15\,km\,s$^{-1}$ tend to produce profiles that are broader than the ones observed. After various tests, we adopted a $v\sin i$ of 11\,km\,s$^{-1}$ for the convolution of our final model.

There are only two \textit{IUE} spectra of HD\,63425. Interestingly, although the optical spectra of HD\,63425 do not show any variation, these UV spectra (obtained over 3 years apart) reveal large variations in the wind lines (see the two spectra shown in Fig. \ref{fig|uv}). The SWP\,21679 spectrum presents these features larger and deeper than the ones found in SWP\,30485. An interesting possibility was proposed to explain a similar issue in the star $\tau$\,Sco. The passage of the magnetic equator across the line-of-sight would reveal material trapped by the magnetic field, providing the extra absorption observed \citep{2006MNRAS.370..629D}. Unfortunately, given the scarcity of UV data for HD\,63425 we have no means to test this idea at present (for example, by verifying the existence of absorption episodes).

We chose to focus on the SWP\,30485 spectrum, where we have shallower absorption profiles. Nevertheless, the mass-loss rate obtained can be safely used as an upper limit.
Our fit is illustrated in Figure \ref{fig|uv}. Again, the use of X-rays was mandatory in producing a N\,\textsc{v} feature comparable to the observed one. We note that the C\,\textsc{iv}\,$\lambda \lambda 1548, 1550$ absorptions could be partially or entirely due to the interstellar medium. We summarize the wind parameters obtained in Table \ref{tab|param}, which are similar to those of HD\,66665.

\section{Magnetic field}

\begin{figure*}
\includegraphics[width=84mm]{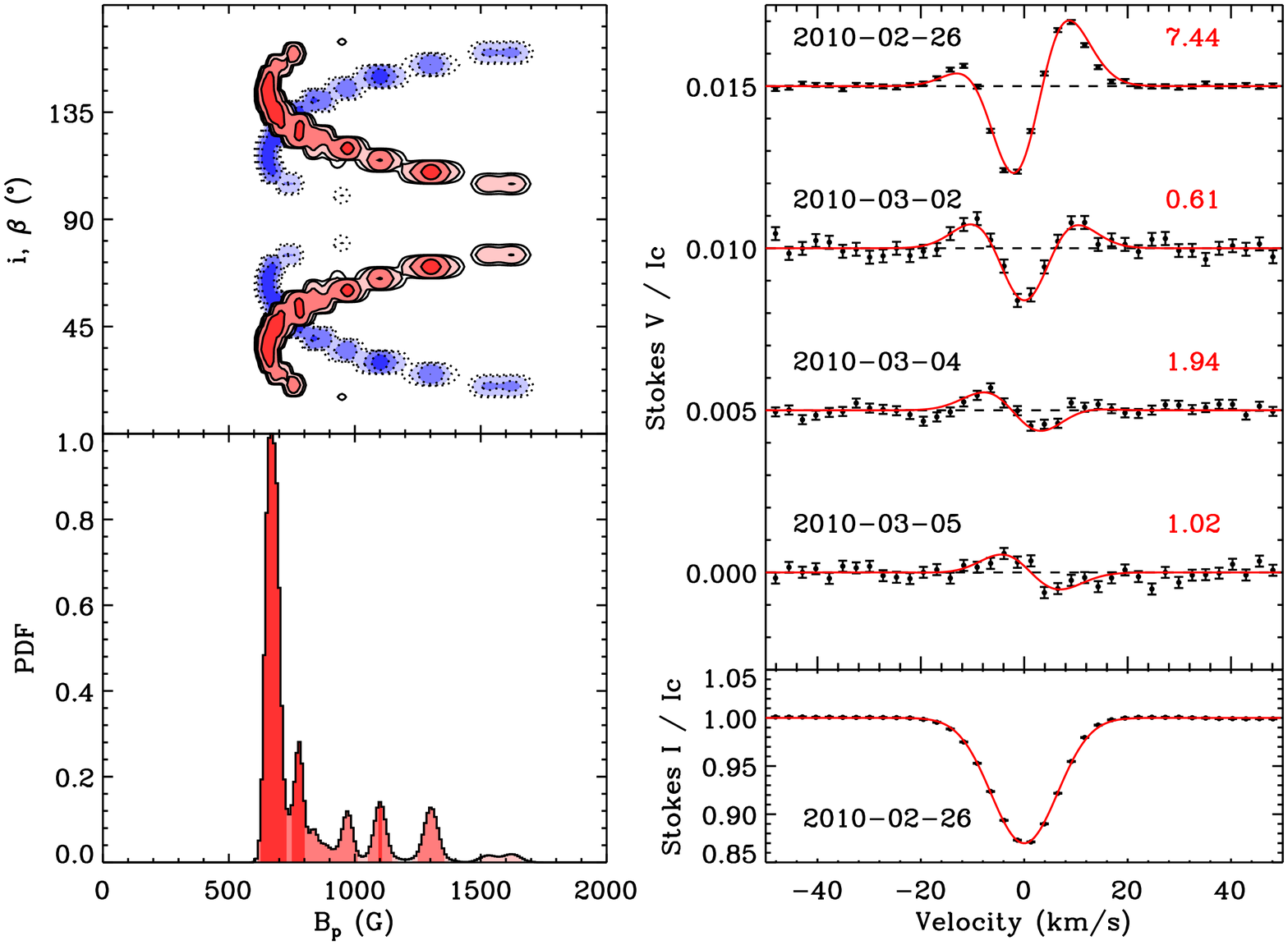}
\includegraphics[width=84mm]{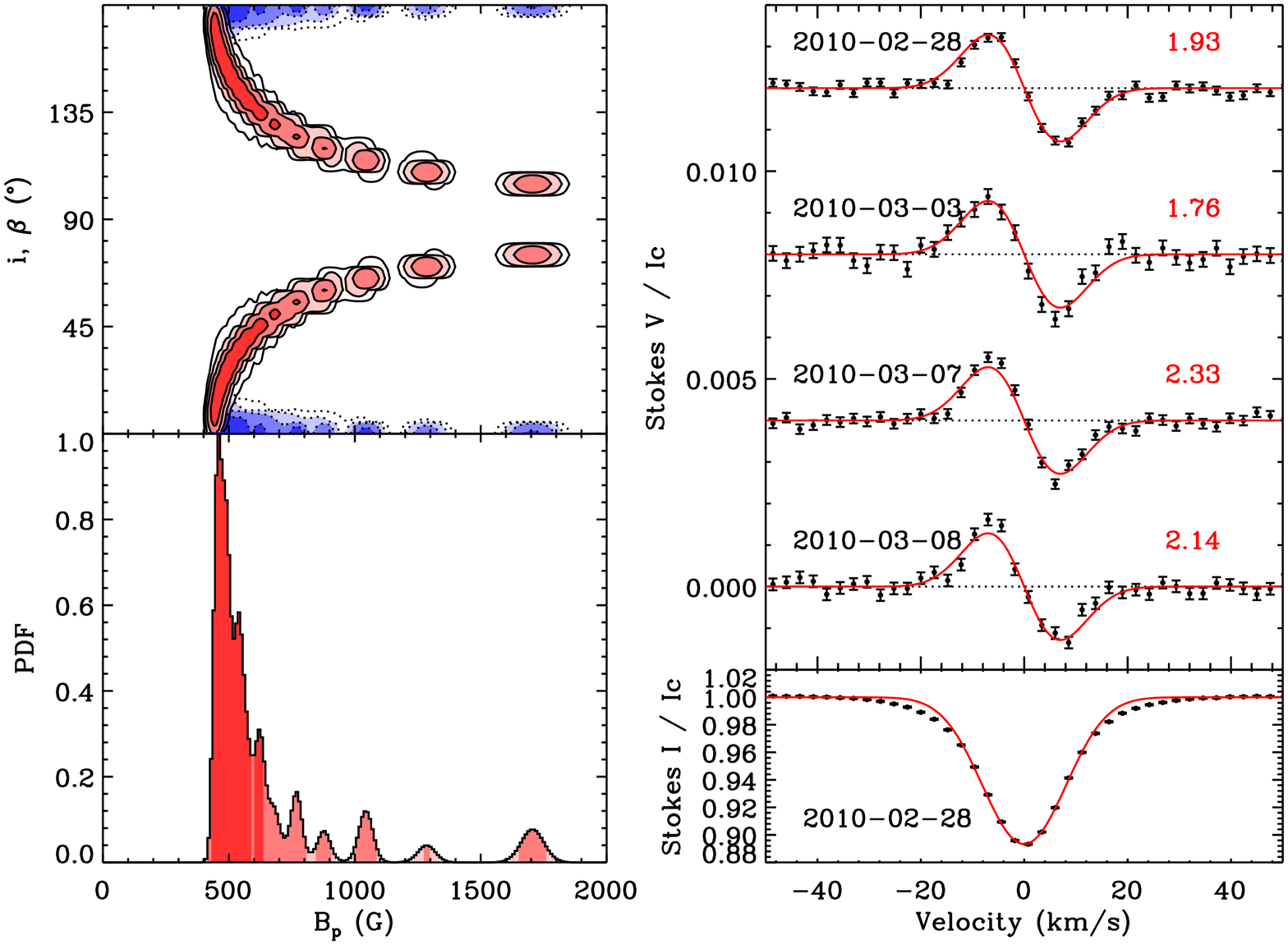}
\caption{\label{fig|mag} Magnetic modelling of HD\,66665 (left) and HD\,63425 (right). For each star, the left panel shows the posterior probability density marginalized for the dipole strength $B_p$ (bottom) and for the 2D $B_p$-$i$ (solid red) and $B_p$-$\beta$ (dashed blue) planes (top). The credible regions containing 68.3\%, 95.4\%, 99\% and 99.7\% of the probability are shaded, from dark to light colours respectively. The right panel shows the LSD Stokes I profile for one observation (bottom) and all LSD Stokes V profiles (top). 
The synthetic profiles produced by the ($B_p$, $i$, $\beta$)-configuration correspond to the peak of the joint posterior probability density, the best phase being extracted for each observation. The corresponding $\chi^2_\mathrm{red}$ is indicated. 
}
\end{figure*}

In order to increase the magnetic sensitivity of our data, we applied the Least-Squares Deconvolution (LSD) procedure, as described by \citet{1997MNRAS.291..658D}. This procedure enables the simultaneous use of many lines present in a spectrum to detect a magnetic field Stokes V signature.
Assumptions, limitations and numerical tests of this method are presented by \citet{2010arXiv1008.5115K}.

We adopted lists of metallic spectral lines suitable for early-B stars from the Vienna Atomic Line Database \citep[VALD,][]{2000BaltA...9..590K}. The depth of each line was adjusted to match the observed spectra of each star. In total, around 300 lines and 260 lines were used for HD\,66665 and HD\,63425 respectively. All LSD profiles were computed on a spectral grid with a velocity bin of 2.6\,km\,s$^{-1}$. 

In order to assess the presence of a magnetic signature in our LSD profiles, we computed the $\chi^2$ probability that the observed deviation of the signal, with respect to $V=0$, is only due to noise (the false alarm probability, or FAP). We consider a signal to be detected when this probability is less than $10^{-5}$, as prescribed by \citet{1997MNRAS.291..658D}. As shown in column 5 of Table \ref{tab|obslog}, all observations led to the detection of a magnetic signal. The same analysis was performed on the diagnostic null ($N$) profiles, where no signals were detected.

The global longitudinal component of the magnetic field was inferred from the first moment of each Stokes V LSD profile, as described by \citet{2000MNRAS.313..851W}. The integration range on each side of the Stokes I centre of gravity was set to $\pm$22\,km\,s$^{-1}$ and $\pm$25\,km\,s$^{-1}$ for HD\,66665 and HD\,63425 respectively.
While the longitudinal field provides a useful statistical measure of the line-of-sight component of the field, we stress that we do not use it as the primary diagnostic of the presence of a magnetic field. This is because a large variety of magnetic configurations can produce a null longitudinal field while still generating a detectable Stokes V profile in the velocity-resolved line profiles \citep[see][]{2010arXiv1010.2248P}.

Figure \ref{fig|mag} (left panel, right frame) shows the LSD profiles of HD\,66665. The Stokes V profiles vary on a timescale of at most a few days, although a period determination is not possible at this point. As the exact rotation phases of our observations are not known, we used the method described by \citet{2008MNRAS.387L..23P}, which compares the observed Stokes V profiles to a large grid of synthetic profiles, described by a dipole oblique rotator model. The model is parametrized by the dipole field strength $B_p$, the rotation axis inclination $i$ with respect to the line of sight, the positive magnetic axis obliquity $\beta$ and the rotational phase $\varphi$.  
Assuming that only $\varphi$ may change between different observations of a given star, the goodness-of-fit of a given rotation-independent ($B_p$, $i$, $\beta$) magnetic configuration can be computed to determine configurations that provides good posterior probabilities for all the observed Stokes V profiles, in a Bayesian statistic framework. 

In Figure \ref{fig|mag} (left panel, right frame), we show the synthetic profiles (in solid red) produced by the ($B_p$, $i$, $\beta$)-configuration that produces the peak of the joint posterior probability. The inclined dipole model can reproduce the current observations in a satisfactory fashion, although the $\chi^2_\mathrm{red}$ remains high for two observations (those with the highest s/n). Whether these deviations from the model come from artefacts of the LSD procedure, systematics introduced by our simple line profile modelling, or a more complex field, remains to be investigated once phase-resolved observations are obtained and a more detailed analysis can be performed. 

By treating any features that cannot be explained by the inclined dipole model as additional Gaussian noise, we can obtain a conservative estimate of the model parameters. For HD 66665, the posterior probability density marginalized for $B_p$ (Fig. \ref{fig|mag}, left panel, left frame) peaks at 670\,G, and provides a strong constraint on the dipole field lower limits (606\,G at 99.7\%). However, the PDF extends to high $B_p$ (1.6\,kG at 99.7\%), as the exact field strength depends strongly on the geometry, as illustrated by the 2D PDFs, marginalized for the $B_p$-$i$ and $B_p$-$\beta$ planes.  

The LSD profiles obtained for HD\,63425 are shown in Figure \ref{fig|mag} (right panel, right frame). No significant variations are observed between the profiles. The red curves show the synthetic profiles produced by the 
($B_p$, $i$, $\beta$) configuration that produces the peak of the joint posterior probability. The inclined dipole model can once again reproduce the current observations in a satisfactory fashion, although some details are not reproduced, as shown by the $\chi^2_\mathrm{red}$ values. Note that our simple modelling of the Stokes I profile was not able to reproduce the broad wings of the intensity line profile. 
The posterior probability density marginalized for $B_p$ (right panel, left frame) peaks at 460\,G, and the 99.7\% credible region extend from 406\,G to 1.8\,kG, depending on the exact geometry. As no assumptions are made about the phases or each observation, the model prefers a dipole field that is roughly aligned with the rotation axis ($\beta\sim0^\circ$), as this configuration is more likely to occur with observations taken at random rotational phases. However, given the low $v\sin i$ and the short time span of our observations, it is possible that all our observations were taken at roughly the same rotational phase. This would suggest that the stellar rotation period is much longer than the period of the observations. Given the resemblance of this star with $\tau$\,Sco, and the observed variability of the UV line profiles, this hypothesis is likely. More observations shall therefore be necessary to constrain the exact geometry of the field.

\section{Conclusion}

We have presented the characteristics of two stars - HD\,66665 and HD\,63425 - which we believe are analogues to the magnetic massive star $\tau$\,Sco for two main reasons:
(i) The UV spectra of these stars are similar to the once-unique spectrum of $\tau$\,Sco. 
(ii) We have shown that these two stars host magnetic fields.

We have shown that all three of these stars have similar fundamental properties. However, the mass-loss rates we estimate for HD\,66665 and HD\,63425 are lower than the value adopted for $\tau$\,Sco by \citet[][$6\times10^{-8}$\,M$_\odot$\,yr$^{-1}$]{2006MNRAS.370..629D} in their analysis of that star. 
However, that mass-loss rate \citep{2005A&A...441..711M} is determined from the optical spectrum of $\tau$\,Sco, and could differ systematically from that determined from UV line profiles -- in particular in the presence of a magnetic field.

The derived mass-loss rates for HD\,66665 and HD\,63425 are sufficiently low that they should be considered as stars presenting the weak wind problem \citep{2005A&A...441..735M,2009A&A...498..837M}, since the expected mass-loss rate is about $10^{-8}$\,M$_\odot$\,yr$^{-1}$ -- about 20 time larger than the inferred rates \citep[following the recipe provided by][]{2000A&A...362..295V}.

Our modelling of the LSD Stokes V profiles assuming an inclined dipole model results in surface field strengths\footnote{The surface-averaged modulus of a dipolar magnetic field is equal to 0.77 times the dipole polar field strength.} that are comparable to, or perharps slightly larger than, the mean surface field strength of $\tau$\,Sco. The current observations can be acceptably reproduced by the dipole model, although more phase-resolved observations are required in order to assess the potential complexity of their magnetic fields, and verify if the wind anomalies are linked to the field complexity.

\section*{Acknowledgments}
GAW  acknowledges support from the Discovery Grants programme of the Natural Science and Engineering Research Council of Canada. We thank the anonymous referee, whose helpful comments led to the improvement of the paper.

\bibliographystyle{mn2e}
\bibliography{tauSco}
\clearpage

\label{lastpage}

\end{document}